\documentstyle[aps,eqsecnum,preprint,floats,epsf,epsfig]{revtex}
\begin{document}
\def\be{\begin{eqnarray}}
\def\en{\end{eqnarray}}
\def\non{\nonumber}
\def\la{\langle}
\def\ra{\rangle}
\def\nc{N_c^{\rm eff}}
\def\vp{\varepsilon}
\def\B{{\cal B}}
\def\up{\uparrow}
\def\dw{\downarrow}
\def\vma{{_{V-A}}}
\def\vpa{{_{V+A}}}
\def\smp{{_{S-P}}}
\def\spp{{_{S+P}}}
\def\J{{J/\psi}}
\def\ov{\overline}
\def\Lqcd{{\Lambda_{\rm QCD}}}
\def\pr{{\sl Phys. Rev.}~}
\def\prl{{\sl Phys. Rev. Lett.}~}
\def\pl{{\sl Phys. Lett.}~}
\def\np{{\sl Nucl. Phys.}~}
\def\zp{{\sl Z. Phys.}~}
\def\lsim{ {\ \lower-1.2pt\vbox{\hbox{\rlap{$<$}\lower5pt\vbox{\hbox{$\sim$}
}}}\ } }
\def\gsim{ {\ \lower-1.2pt\vbox{\hbox{\rlap{$>$}\lower5pt\vbox{\hbox{$\sim$}
}}}\ } }

\font\el=cmbx10 scaled \magstep2{\obeylines\hfill October, 2002}

\vskip 1.5 cm

\centerline{\large\bf Hadronic $B$ Decays to Charmed Baryons}
\bigskip
\centerline{\bf Hai-Yang Cheng$^{1}$ and Kwei-Chou Yang$^{2}$}
\medskip
\centerline{$^1$ Institute of Physics, Academia Sinica}
\centerline{Taipei, Taiwan 115, Republic of China}
\medskip

\medskip
\centerline{$^2$ Department of Physics, Chung Yuan Christian
University} \centerline{Chung-Li, Taiwan 320, Republic of China}
\bigskip
\bigskip
\centerline{\bf Abstract}
\bigskip
{\small
 We study exclusive $B$ decays to final states containing a
charmed baryon within the pole model framework. Since the strong
coupling for $\Lambda_b\ov B N$ is larger than that for $\Sigma_b
\ov BN$, the two-body charmful decay $B^-\to\Sigma_c^0\bar p$ has
a rate larger than $\ov B^0\to\Lambda_c^+\bar p$ as the former
proceeds via the $\Lambda_b$ pole while the latter via the
$\Sigma_b$ pole. By the same token, the three-body decay $\ov
B^0\to\Sigma_c^{++}\bar p\pi^-$ receives less baryon-pole
contribution than $B^-\to\Lambda_c^+\bar p\pi^-$. However, because
the important charmed-meson pole diagrams contribute
constructively to the former and destructively to the latter,
$\Sigma_c^{++}\bar p\pi^-$ has a rate slightly larger than
$\Lambda_c^+\bar p\pi^-$. It is found that one quarter of the
$B^-\to \Lambda_c^+\bar p\pi^-$ rate comes from the resonant
contributions. We discuss the decays $\ov B^0\to\Sigma_c^0\bar
p\pi^+$ and $B^-\to\Sigma_c^0\bar p\pi^0$ and stress that they are
not color suppressed even though they can only proceed via an
internal $W$ emission.

}

\pagebreak

\section{Introduction}
Previously CLEO has searched for charmful baryonic $B$ decays in
the class $\ov B\to\Lambda_c\ov N X$. The experimental results are
\cite{CLEO97}:
 \be \label{data}
 \B(\ov B^0\to\Lambda_c^+\bar p\pi^+\pi^-) &=& (1.33^{+0.46}_{-0.40}\pm0.37)\times
 10^{-3}, \non \\
 \B(B^-\to\Lambda_c^+\bar p\pi^-\pi^0)&<& 3.12\times
 10^{-3},  \\
  \B(B^-\to\Lambda_c^+\bar p\pi^-) &=& (6.2^{+2.3}_{-2.0}\pm1.6)\times 10^{-4},  \non \\
 \B(\ov B^0\to \Lambda_c^+\bar p)&<& 2.1\times
 10^{-4}. \non
 \en
Recently, Belle \cite{Bellebaryon2} and CLEO \cite{CLEO02} have
reported the measurements of the exclusive decays of $B$ mesons
into final states of the type $\B_c\bar pn(\pi)$, where
$\B_c=\Lambda_c,\Lambda_{c1},\Sigma_c(2455),\Sigma_{c1}$
[$\Lambda_{c1}=\Lambda_c(2593),\Lambda_c(2625)$ and
$\Sigma_{c1}=\Sigma_c(2520)$] and $n$ is the number of the pions
in the final state. Form Table I we see that the new measurements
of $B^-\to\Lambda_c^+\bar p\pi^-$ and $\ov B^0\to\Lambda_c^+\bar
p\pi^+\pi^-$ are consistent with, and much more accurate than, the
previous CLEO results (\ref{data}), however the new result for the
former is somewhat low $(1.5\sigma$).

\begin{table}[ht]
\caption{Experimental measurements of the branching ratios (in
units of $10^{-4}$) for the $B$ decay modes with a charmed baryon
$\Lambda_c$ or $\Lambda_{c1}=\Lambda_c(2593),\Lambda_c(2625)$ or
$\Sigma_c(2455)$ or $\Sigma_{c1}=\Sigma_c(2520)$ in the final
state.
 }
\footnotesize
\begin{center}
\begin{tabular}{l l l   }
 Mode~~ & Belle \cite{Bellebaryon2} & CLEO  \cite{CLEO02} \\ \hline
  $B^-\to\Lambda_c^+\bar p\pi^-\pi^0$ & &
 $18.1\pm2.9^{+2.2}_{-1.6}\pm4.7$ \\
 $\ov B^0\to\Lambda_c^+\bar p\pi^+\pi^-$ &
 $11.0\pm1.2\pm1.9\pm 2.9$ &
 $16.7\pm1.9^{+1.9}_{-1.6}\pm4.3$  \\
 $B^-\to \Lambda_c^+\bar p\pi^-$ & $1.87^{+0.43}_{-0.40}\pm0.28\pm
 0.49$ & $2.4\pm0.6^{+0.19}_{-0.17}\pm0.6$  \\
 $\ov B^0\to\Lambda_c^+\bar p$ &
 $0.12^{+0.10}_{-0.07}\pm0.02\pm0.03<0.31$ & $<0.9$  \\ \hline
 $B^-\to\Lambda_{c1}^+\bar p\pi^-$ & & $<1.9$ \\
 $\ov B^0\to\Lambda_{c1}^+\bar p$ & & $<1.1$ \\ \hline
 $B^-\to \Sigma_c^{++}\bar p\pi^-\pi^-$ & & $2.8\pm 0.9\pm 0.5\pm
 0.7$  \\
 $B^-\to\Sigma_c^0\bar p\pi^+\pi^-$ & & $4.4\pm1.2\pm0.5\pm1.1$ \\
 $\ov B^0\to \Sigma_c^{++}\bar p\pi^-$ &
 $2.38^{+0.63}_{-0.55}\pm0.41\pm0.62$ & $3.7\pm0.8\pm0.7\pm0.8$
 \\
 $\ov B^0\to\Sigma_c^0\bar p\pi^+$ &
 $0.84^{+0.42}_{-0.35}\pm0.14\pm0.22<1.59$ & $ 2.2\pm0.6\pm0.4\pm0.5$
 \\
 $B^-\to\Sigma_c^0\bar p\pi^0$ & & $4.2\pm 1.3\pm0.4\pm1.0$ \\
 $B^-\to\Sigma_c^0\bar p$ & $0.45^{+0.26}_{-0.19}\pm0.07\pm0.12<0.93$ &
 $<0.8$  \\ \hline
 $\ov B^0\to \Sigma_{c1}^{++}\bar p\pi^-$ &
 $1.63^{+0.57}_{-0.51}\pm0.28\pm0.42$ & \\
 $\ov B^0\to\Sigma_{c1}^0\bar p\pi^+$ &
 $0.48^{+0.45}_{-0.40}\pm0.08\pm0.12<1.21$ &  \\
 $B^-\to\Sigma_{c1}^0\bar p$ & $0.14^{+0.15}_{-0.09}\pm0.02\pm0.04<0.46$ &  \\
\end{tabular}
\end{center}
\end{table}

In general, CLEO and Belle results are consistent with each other
except for the ratio of $\Sigma_c^{++}\bar p\pi^-$ to
$\Sigma_c^0\bar p\pi^+$. The $\Sigma_c^{++}$ decay proceeds via
both external and internal $W$-emission diagrams, whereas the
$\Sigma_c^0$ decay can only proceed via an internal $W$ emission.
While Belle measurements imply a sizable suppression for the
$\Sigma_c^0$ decay (and likewise for the $\Sigma_{c1}$ decay), it
is found by CLEO that $\Sigma_c^{++}\bar p\pi^-$, $\Sigma_c^0\bar
p\pi^+$ and $\Sigma_c^0\bar p\pi^0$ are of the same order of
magnitude. Therefore, it is concluded by CLEO that the external
$W$ decay diagram does not dominate over the internal $W$-emission
diagram in Cabibbo-allowed baryonic $B$ decays. This needs to be
clarified by the forthcoming improved measurements.

On the theoretical side, the decays $\ov B^0\to\Lambda_c^+\bar p$
and $B^-\to\Lambda_c^+\bar p\pi^-$ have been studied by us within
the framework of the pole model \cite{CYcharmful}. We have
explained several reasons why the three-body decay rate of $B^-\to
\Lambda_c^+\bar p\pi^-$ is larger than that of the two-body one
$\ov B^0\to \Lambda_c^+\bar p$.  At the pole-diagram level, the
$\Sigma_b$ propagator in the pole amplitude for the latter is of
order $1/(m_b^2-m_c^2)$, while the invariant mass of the
$(\Lambda_c^+\pi^-)$ system can be large enough in the former
decay so that its propagator of $\Lambda_b$ in the pole diagram is
not subject to the same $1/m_b^2$ suppression. Moreover, the
strong coupling constant for $\Lambda_b^0\to B^- p$ is larger than
that for $\Sigma_b^+\to\ov B^0 p$, and this suffice to explain the
original CLEO observation.

Since at the pole-diagram level, $\ov B^0\to\Sigma_c^{++}\bar
p\pi^-$ proceeds through the $\Sigma_b$ pole, while
$B^-\to\Sigma_c^0\bar p$ proceeds through the $\Lambda_b$ pole, it
is naively expected that $\Gamma(B^-\to\Sigma_c^0\bar
p)>\Gamma(\ov B^0\to\Lambda_c^+\bar p)$ and $\Gamma(\ov
B^0\to\Sigma_c^{++}\bar p\pi^-)<\Gamma(B^-\to\Lambda_c^+\bar
p\pi^-)$. However, the latter relation is not borne out by the new
measurements of both Belle and CLEO (see Table I). Indeed, at the
quark level, it appears that $\Sigma_c^{++}\bar p\pi^-$ and
$\Lambda_c^+\bar p\pi^-$ should have similar rates as both of them
receive external $W$-emission contributions.

It turns out that the meson-pole contribution to the three-body
baryonic $B$ decays which was originally missed in
\cite{CYcharmful} is important for the charmful $B$ decays $\ov
B^0\to\Sigma_c^{++}\bar p\pi^-$ and $B^-\to\Lambda_c^+\bar
p\pi^-$. Moreover, this meson-pole effect contributes
destructively to $\Lambda_c^+\bar p\pi^-$ and constructively to
$\Sigma_c^{++}\bar p\pi^-$. As we shall see, this eventually leads
to the explanation of why $\B(\ov B^0\to\Sigma_c^{++}\bar
p\pi^-)\gsim \B(B^-\to\Lambda_c^+\bar p\pi^-)$.

Since $\ov B^0\to\Sigma_c^0\bar p\pi^+$ and $B^-\to\Sigma_c^0\bar
p\pi^0$ can only proceed via an internal $W$ emission, it is
suitable to apply the pole model to study these two decays. As we
shall see later, not all the internal $W$-emission diagrams in
baryonic decays are subject to color suppression.

The layout of the present paper is organized as follows. In Sec.
II we first study the two-body charmful decay $\ov B^0\to
\Lambda_c^+ \,\bar p$, $B^-\to\Sigma_c^0\bar p$, $\ov
B^0\to\Sigma_c^0\bar n$ and $B^-\to\Lambda_c^+\bar\Delta^{--}$. We
then turn to the three-body decays $\ov B^0\to \Lambda_c^+ \,\bar
p\,\pi^-$, $\ov B^0\to\Sigma_c^{++}\bar p\pi^-$, $\ov
B^0\to\Sigma_c^0\bar p\pi^+$ and $B^-\to\Sigma_c^0\bar p\pi^0$ in
Sec. III. Discussions and conclusions are given in Sec. IV.

\section{Two-body charmful $B$ Decays}
In this section we shall study the two-body charmful decays $\ov
B^0\to \Lambda_c^+ \,\bar p$, $B^-\to\Sigma_c^0\bar p$, $\ov
B^0\to\Sigma_c^0\bar n$ and $B^-\to\Lambda_c^+\bar\Delta^{--}$.
Since the former has been discussed in \cite{CYcharmful}, we will
describe it in a somewhat cursory way.

\subsection{$\ov B^0\to\Lambda_c^+\bar p$}
To proceed, we first write down the Hamiltonian relevant for the
present paper
 \be \label{Ham}
 {\cal H}_{\rm eff} &=& {G_F\over\sqrt{2}}
 V_{cb}V_{ud}^*[c_1^{\rm eff}O_1+c_2^{\rm eff}O_2]+H.c.,
 \en
where $O_1=(\bar cb)(\bar d u)$ and $O_2=(\bar cu)(\bar db)$ with
$(\bar q_1q_2)\equiv \bar q_1\gamma_\mu(1-\gamma_5)q_2$ and the
effective coefficients $c_1^{\rm eff}$ and $c_2^{\rm eff}$ are
renormalization scale and scheme independent. In order to ensure
that the physical amplitude is renormalization scale and
$\gamma_5$-scheme independent, we have included vertex corrections
to the hadronic matrix elements. This amounts to redefining the
Wilson coefficients $c_{1,2}(\mu)$ into the effective ones
$c_{1,2}^{\rm eff}$. Numerically we have $c_1^{\rm eff}=1.168$ and
$c_2^{\rm eff}=-0.365$ \cite{CCTY}.

The decay amplitude of $\ov B^0\to\Lambda_c^+\,\bar p$ consists of
factorizable and nonfactorizable parts:
 \be
 A(\ov B^0\to\Lambda_c^+\,\bar p)=A(\ov B^0\to\Lambda_c^+\,\bar p)_{\rm fact}
 +A(\ov B^0\to\Lambda_c^+\,\bar p)_{\rm nonfact},
 \en
with
 \be
 A(\ov B^0\to\Lambda_c^+\,\bar p)_{\rm
 fact}={G_F\over\sqrt{2}}V_{cb}V_{ud}^*\,a_2\la \Lambda_c \bar p|(\bar
 cu)|0\ra\la 0|(\bar d b)|\ov B^0\ra,
 \en
where $a_2=c_2^{\rm eff}+c_1^{\rm eff}/N_c$. The short-distance
factorizable contribution is nothing but the $W$-exchange diagram.
This $W$-exchange contribution has been estimated and is found to
be very small and hence can be neglected \cite{Korner,Kaur}.
However, a direct evaluation of nonfactorizable contributions is
very difficult.  It is customary to assume that the
nonfactorizable effect is dominated by the pole diagram with
low-lying baryon intermediate states; that is, nonfactorizable
$s$- and $p$-wave amplitudes are dominated by ${1\over 2}^-$
low-lying baryon resonances and ${1\over 2}^+$ ground-state
intermediate states, respectively \cite{Jarfi}. For $\ov
B^0\to\Lambda_c^+\,\bar p$, we consider the strong-interaction
process $\ov B^0\to \Sigma_b^{+(*)}\,\bar p$ followed by the weak
transition $\Sigma_b^{+(*)}\to \Lambda_c$, where $\Sigma_b^*$ is a
${1\over 2}^-$ baryon resonance (see Fig. 1). The pole-diagram
amplitude has the form
 \be
 A(\ov B^0\to\Lambda_c^+\bar p)_{\rm nonfact}=\bar u_{\Lambda_c}(A+B\gamma_5)v_{\bar p},
 \en
where
 \be
 A=-{g_{\Sigma_b^{+*}\to \ov B^0p}b_{\Sigma_b^{+*}\Lambda_c^+}\over
 m_{\Lambda_c}-m_{\Sigma_b^*} }, \qquad\qquad B=\,{g_{\Sigma_b^+\to\ov B^0p}
 a_{\Sigma_b^+\Lambda_c^+}\over m_{\Lambda_c}-m_{\Sigma_b}},
 \en
correspond to $s$-wave parity-violating (PV) and $p$-wave
parity-conserving (PC) amplitudes, respectively, and
 \be
 \la \Lambda_c^+|{\cal H}_{\rm eff}^{\rm PC}|\Sigma_b^+\ra = \bar
 u_{\Lambda_c}a_{\Sigma_b^+\Lambda_c^+}u_{\Sigma_b}, \qquad\quad
 \la \Lambda_c^+|{\cal H}_{\rm eff}^{\rm PV}|\Sigma_b^{*+}\ra = i\bar
 u_{\Lambda_c}b_{\Sigma_b^{+*}\Lambda_c^+}u_{\Sigma_b^*}.
 \en

\begin{figure}[tb]
\hspace{2cm}\psfig{figure=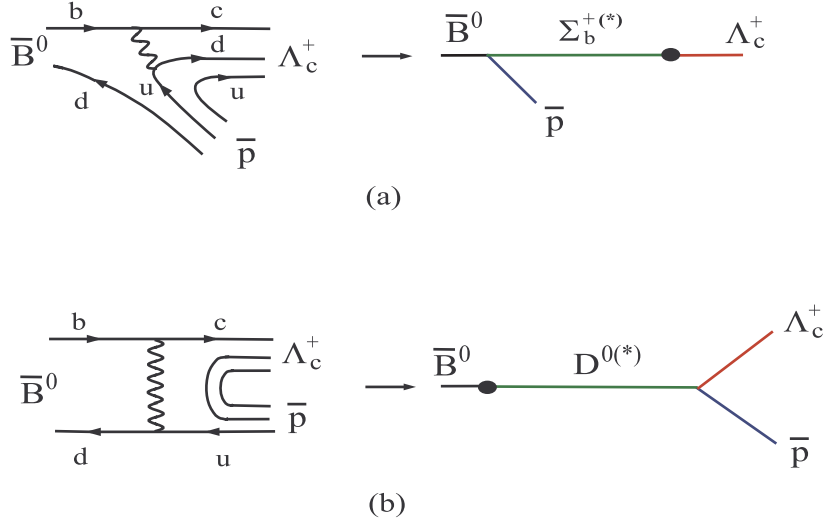,height=3in} \vspace{1.2cm}
    \caption{{\small Quark and pole diagrams for $\ov B^0\to\Lambda_c^+\bar
    p$, where the solid blob denotes the weak vertex. Fig. 1(a) corresponds
    to a nonfactorizable internal
    $W$ emission, while Fig. 1(b) to a $W$-exchange contribution.
    }}
   \label{fig:1}
\end{figure}

The main task is to evaluate the weak matrix elements and the
strong coupling constants. We shall employ the MIT bag model
\cite{MIT} to evaluate the baryon matrix elements (see e.g.
\cite{CT92,CT93} for the method). Since the quark-model wave
functions best resemble the hadronic states in the frame where
both baryons are static, we thus adopt the static bag
approximation for the calculation. Note that because the
combination of the four-quark operators $O_1+O_2$ is symmetric in
color indices, it does not contribute to the baryon-baryon matrix
element since the baryon-color wave function is totally
antisymmetric. This leads to the relation
$\la\Lambda_c^+|O_2|\Sigma_b^+\ra=-\la\Lambda_c^+|O_1|\Sigma_b^+\ra$.
From Eq. (\ref{Ham}) we obtain the PC matrix element
 \be \label{PCm.e.}
 a_{\Sigma_b^+\Lambda_c^+}=-{G_F\over
 \sqrt{2}}V_{cb}V_{ud}^*\,(c_1^{\rm eff}-c_2^{\rm eff}){2\over\sqrt{6}}
 (X_1+3X_2)(4\pi),
 \en
where\footnote{For details of the MIT bag model evaluation, see
\cite{CYcharmful,CYBaryon}. Note that  the bag integrals $X_1$ and
$X_2$ given in Eq. (B4) of \cite{CYBaryon} are defined for the
operator $O_2$ rather than for $O_1$. }
 \be \label{bagX}
 X_1 &=& \int^R_0
 r^2dr[u_d(r)v_b(r)-v_d(r)u_b(r)][u_c(r)v_u(r)-v_c(r)u_u(r)], \non
 \\
 X_2 &=& \int^R_0
 r^2dr[u_d(r)u_b(r)+v_d(r)v_b(r)][u_c(r)u_u(r)+v_c(r)v_u(r)]
 \en
are four-quark overlap bag integrals and $u_q(r)$, $v_q(r)$ are
the large and small components of the quark wave functions in the
ground $(1S_{1/2})$ state. In principle, one can also follow
\cite{CT92} to tackle the low-lying negative-parity $\Sigma_b^*$
state in the bag model and evaluate the PV matrix element
$b_{\Sigma^*_c\Lambda_c}$. However, it is known that the bag model
is less successful even for the physical non-charm and non-bottom
${1\over 2}^-$ resonances \cite{MIT}, not mentioning the charm or
bottom ${1\over 2}^-$ resonances. In short, we know very little
about the ${1\over 2}^-$ state. Therefore, we will not evaluate
the PV matrix element $b_{\Sigma_b^*\Lambda_c}$ as its calculation
in the bag model is much involved  and is far more uncertain than
the PC one \cite{CT92}.

Using the bag wave functions given in the Appendix of
\cite{CYcharmful}, we find numerically
 \be
 X_1=-1.49\times 10^{-5}\,{\rm GeV}^3, \qquad\quad
 X_2=1.81\times 10^{-4}\,{\rm GeV}^3.
 \en
The decay rate of $B\to\B_1\ov \B_2$ is given by
 \be
 \Gamma(B\to \B_1\ov \B_2)&=& {p_c\over 4\pi}\Bigg\{
 |A|^2\,{(m_B+m_1+m_2)^2p_c^2\over (E_1+m_1)(E_2+m_2)m_B^2}\non \\
 & +&
 |B|^2\,{[(E_1+m_1)(E_2+m_2)+p_c^2]^2\over
 (E_1+m_1)(E_2+m_2)m_B^2} \Bigg\},
 \en
where $p_c$ is the c.m. momentum, $E_i$ and $m_i$ are the energy
and mass of the baryon $\B_i$, respectively. Putting everything
together we obtain
 \be
 \B(\ov B^0\to\Lambda_c^+\bar p)_{\rm PC}=5.0\times
 10^{-6}\left|{g_{\Sigma_b^+\to \ov B^0p}\over 5}\right|^2.
 \en
The PV contribution is expected to be smaller. For example, it is
found to be $\Gamma^{\rm PV}/\Gamma^{\rm PC}=0.59$ in
\cite{Jarfi}. Therefore, we conclude that
 \be \label{Lamcp}
 \B(\ov B^0\to\Lambda_c^+\bar p)\lsim 7.9\times
 10^{-6}\left|{g_{\Sigma_b^+\to \ov B^0p}\over 5}\right|^2.
 \en
The strong coupling $g_{\Sigma_b^+\to \ov B^0p}$ has been
estimated in \cite{Jarfi} using the $^3P_0$ quark-pair-creation
model and it is found to lie in the range $|g_{\Sigma_b^+\to\ov
B^0p}|=6\sim 10$.  At any rate, the prediction (\ref{Lamcp}) is
consistent with the current experimental limit of $3.1\times
10^{-5}$ by Belle \cite{Bellebaryon2} and $9\times 10^{-5}$ by
CLEO \cite{CLEO02}. Note that all earlier predictions based on the
QCD sum rule \cite{Chernyak} or the pole model \cite{Jarfi} or the
diquark model \cite{Ball} are too large compared to experiment
(see e.g. Table I of \cite{CYcharmful}). In the pole-model
calculation in \cite{Jarfi}, the weak matrix element is largely
over-estimated.

\subsection{$B^-\to\Sigma_c^0\bar p$ and $\ov B^0\to\Sigma_c^0\bar n$}
The pole diagrams for $B^-\to\Sigma_c^0\bar p$ and $\ov
B^0\to\Sigma_c^0\bar n$ consist of two poles: $\Lambda_b^{0(*)}$
and $\Sigma_b^{0(*)}$ as depicted in Fig. 2. Proceeding as before,
the parity-conserving amplitudes read\footnote{It is found in
\cite{Jarfi} that the parity-violating contribution to
$B^-\to\Sigma_c^0\bar p$ is largely suppressed relative to the
parity-conserving one.}
 \be \label{PCSigcp}
 B(B^-\to\Sigma_c^0\bar p) &=& {g_{\Lambda_b^0\to
 B^-p}\,a_{\Lambda_b^0\Sigma_c^0}\over m_{\Sigma_c}-m_{\Lambda_b}}+{g_{\Sigma_b^0\to
 B^-p}\,a_{\Sigma_b^0\Sigma_c^0}\over m_{\Sigma_c}-m_{\Sigma_b}},
 \non \\
 B(\ov B^0\to\Sigma_c^0\bar n) &=& {g_{\Lambda_b^0\to
 \ov B^0n}\,a_{\Lambda_b^0\Sigma_c^0}\over m_{\Sigma_c}-m_{\Lambda_b}}+{g_{\Sigma_b^0\to
 \ov B^0n}\,a_{\Sigma_b^0\Sigma_c^0}\over m_{\Sigma_c}-m_{\Sigma_b}},
 \en
where
  \be \label{aLambSigc}
 a_{\Lambda_b^0\Sigma_c^0} &=& -{G_F\over
 \sqrt{2}}V_{cb}V_{ud}^*\,(c_1^{\rm eff}-c_2^{\rm eff}){2\over\sqrt{6}}
 (X_1-3X_2)(4\pi), \non \\
 a_{\Sigma_b^0\Sigma_c^0} &=& {G_F\over
 \sqrt{2}}V_{cb}V_{ud}^*\,(c_1^{\rm eff}-c_2^{\rm eff}){\sqrt{2}\over 3}
 (X_1+9X_2)(4\pi).
 \en

There are two models which can be used to estimate the strong
couplings: the $^3P_0$ quark-pair-creation model in which the
$q\bar q$ pair is created from the vacuum with vacuum quantum
numbers $^3P_0$, and the $^3S_1$ model in which the quark pair is
created perturbatively via one gluon exchange with one-gluon
quantum numbers $^3S_1$.  Presumably, the $^3P_0$ model works in
the nonperturbative low energy regime. In contrast, in the
perturbative high energy region where perturbative QCD is
applicable, it is expected  the $^3S_1$ model may be more relevant
as the light baryons produced in two-body charmless baryonic $B$
decays are very energetic. However, in practice it is much simpler
to estimate the relative strong coupling strength in the $^3P_0$
model  \cite{Jarfi,Yaouanc} rather than in the $^3S_1$ model where
hard gluons arise from four different quark legs and generally
involve infrared problems.

\begin{figure}[tb]
\hspace{2cm}\psfig{figure=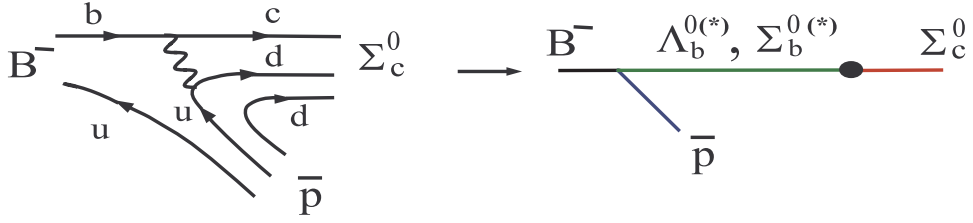,height=1.5in}
\vspace{1.2cm}
    \caption{{\small The internal $W$-emission diagram and its corresponding
    pole diagram for $B^-\to\Sigma_c^0\bar
    p$, where the solid blob denotes the weak vertex.
    }}
\end{figure}

In the $^3P_0$ model we have the relations (see Eq. (3.23) of
\cite{CYBaryon})
 \be \label{3P0}
 g_{\Lambda_b^0\to B^-p} &=& 3\sqrt{3}\,g_{\Sigma_b^0\to B^-p}=-3\sqrt{3\over 2}
 \,g_{\Sigma_b^+\to \ov B^0p}\,, \non \\
  g_{\Lambda_b^0\to \ov B^0n}&=& -3\sqrt{3}\,g_{\Sigma_b^0\to \ov B^0n}=3\sqrt{3\over 2}
 \,g_{\Sigma_b^+\to \ov B^0p}\,.
 \en
This leads to $|g_{\Lambda_b^0\to B^-p}|=18$ for
$|g_{\Sigma_b^+\to \ov B^0p}|=5$. However, the predicted branching
ratio $1.6\times 10^{-4}$ for $B^-\to\Sigma_c^0\bar p$ is too
large compared to the experimental limits, $0.93\times 10^{-4}$ by
Belle and $0.8\times 10^{-4}$ by CLEO. This means that the $^3P_0$
model relation Eq. (\ref{3P0}) is badly broken. This is not a
surprise: As discussed above, the relevant model for energetic
two-body baryonic $B$ decays is the $^3S_1$ model. Fitting to the
central value of the measured branching ratio of
$B^-\to\Sigma_c^0\bar p$, $0.45\times 10^{-4}$ (see Table I), we
find
 \be \label{3P0m}
 g_{\Lambda_b^0\to B^-p}\approx 1.2\sqrt{3}\,g_{\Sigma_b^0\to B^-p}=-1.2\sqrt{3\over 2}
 \,g_{\Sigma_b^+\to \ov B^0p}.
 \en
The isospin relation leads to
  \be
g_{\Lambda_b^0\to \ov B^0n}\approx -1.2\sqrt{3}\,g_{\Sigma_b^0\to
\ov B^0n}=1.2\sqrt{3\over 2} \,g_{\Sigma_b^+\to \ov B^0p}.
 \en
Hence,  $|g_{\Lambda_b^0\to B^-p}|=|g_{\Lambda_b^0\to \ov
B^0n}|\sim 7$ and $|g_{\Sigma_b^0\to B^-p}|=|g_{\Sigma_b^0\to \ov
B^0n}|\sim 3.5$ for $|g_{\Sigma_b^+\to \ov B^0p}|=5$. Since
$|g_{\Lambda_b^0\to B^-p}|>|g_{\Sigma_b^0\to B^-p}|$, $B^-\to
\Sigma_c^0\bar p$ has a larger rate than $\ov
B^0\to\Lambda_c^+\bar p$. Note that $B^-\to\Sigma_c^0\bar p$ is
thus far the only two-body baryonic $B$ decay that its evidence
has been observed by Belle with a significance of $3\sigma$
\cite{Bellebaryon2}.

In contrast, the decay rate of $\Sigma_c^0\bar n$ is quite
suppressed,
 \be
 \B(\ov B^0\to\Sigma_c^0\bar n)=6\times 10^{-7}.
 \en
It has something to do with the smallness of the weak transition
for $\ov B^0\to\Sigma_c^0\bar n$. Since $X_1\ll X_2$, to a good
approximation we have $a_{\Lambda_b^0\Sigma_c^0}\approx
a_{\Sigma_b^0\Sigma_c^0}/\sqrt{3}$ [see Eq. (\ref{aLambSigc})]. As
$g_{\Lambda_b^0\to \ov B^0n}\approx -1.2\sqrt{3}\,g_{\Sigma_b^0\to
\ov B^0n}$, there is a large cancellation occurred in the PC
amplitude, see Eq. (\ref{PCSigcp}). Note that the ratio $R\equiv
\Gamma(\ov B^0\to\Sigma_c^0\bar n)/\Gamma(B^-\to\Sigma_c^0\bar p)$
is predicted to be 1/2 in the $^3P_0$ model \cite{Jarfi}, whereas
it is only of order $10^{-2}$ in our case. Therefore, a
measurement of the ratio $R$ can be used to discriminate between
different quark-pair-creation models.

It should be stressed again that the strong couplings are in
principle $q^2$ dependent. Therefore, the values of strong
couplings quoted above should be considered as an average over the
allowed $q^2$ region.

\subsection{$B^-\to\Lambda_c^+\bar\Delta^{--}$}
The relevant pole diagram for the decay
$B^-\to\Lambda_c^+\bar\Delta^{--}$ ($\bar\Delta^{--}$ being the
antiparticle of $\Delta^{++}$) consists of the intermediate states
$\Sigma_b^{+(*)}$. Since the parity-violating amplitude vanishes
in the $^3P_0$ quark-pair-creation model \cite{Korner,Jarfi}, we
thus have
 \be
 C=0, \qquad\qquad  D=\,{g_{\Sigma_b^+\to
 B^-\Delta^{++}}\,a_{\Sigma_b^+\Lambda_c^+}\over
 m_{\Lambda_c}-m_{\Sigma_b}},
 \en
corresponding to the parity-violating $p$-wave and
parity-conserving $d$-wave amplitudes, respectively, for the decay
$\ov B\to \B_1({1\over 2}^+)\ov \B_2({3\over 2}^-)$ with a
spin-${3\over 2}$ baryon in the final state,
 \be
 {\cal A}(\ov B\to \B_1(p_1)\ov \B_2(p_2))=iq_\mu\bar u_1(p_1)(C+D\gamma_5)v^\mu_2(p_2),
 \en
where $v^\mu$ is the Rarita-Schwinger vector spinor for a
spin-${3\over 2}$ antiparticle and $q=p_1-p_2$. The  decay rate is
 \be
 \Gamma(\ov B\to \B_1({1/ 2}^+)\ov \B_2({3/ 2}^-)) &=& {p_c^3\over 6\pi}\,{1\over m_1^2}\Bigg\{
 |C|^2\,{[(E_1+m_1)(E_2+m_2)+p_c^2]^2\over
 (E_1+m_1)(E_2+m_2)m_B^2} \non \\
 & +& |D|^2\,{(m_B+m_1+m_2)^2p_c^2\over (E_1+m_1)(E_2+m_2)m_B^2}
  \Bigg\}.
 \en

In the $^3P_0$ model one has the relation (see e.g. Eq. (3.32) of
\cite{CYBaryon})
 \be \label{gDelta}
 g_{\Sigma_b^+\to B^-\Delta^{++}} = 2\sqrt{6}\,g_{\Sigma_b^+\to\ov
 B^0p}\,.
 \en
As before, this $^3P_0$ model relation is also expected to be
badly broken. Indeed, it has been pointed out in \cite{CYBaryon}
that using the strong coupling $g_{\Sigma_b^+\to B^-\Delta^{++}}$
extracted from Eq. (\ref{gDelta}) will lead to $\B(B^-\to
p\bar\Delta^{--})=5.8\times 10^{-6}$. Because of the strong decay
$\bar\Delta^{--}\to\bar p\pi^-$, the resonant contribution from
$\bar\Delta^{--}$ to the branching ratio of $B^-\to p\bar p\pi^-$
would be $6\times 10^{-6}$. This already exceeds the recent Belle
measurement $\B(B^-\to p\bar
p\pi^-)=(1.9^{+1.0}_{-0.9}\pm0.3)\times 10^{-6}$ or the upper
limit of $3.7\times 10^{-6}$ \cite{Bellebaryon}. Therefore, the
coupling of the $\Delta$ to the $B$ meson and the octet baryon is
smaller than what is expected from Eq. (\ref{gDelta}). By applying
the same scaling from Eq. (\ref{3P0}) to Eq. (\ref{3P0m}), it is
natural to have
 \be
 g_{\Sigma_b^+\to B^-\Delta^{++}} \approx 0.8\sqrt{6}\,g_{\Sigma_b^+\to\ov
 B^0p}\,.
 \en
Therefore,  $g_{\Sigma_b^+\to B^-\Delta^{++}}=9.8$ for
$g_{\Sigma_b^+\to\ov B^0p}=5$, which is close to the value of 12
employed in \cite{CYBaryon}. Numerically, we obtain
 \be
 \B(B^-\to\Lambda_c^+\bar\Delta^{--})=1.9\times 10^{-5},
 \en
where use of Eq. (\ref{PCm.e.}) has been made.

\section{Three-body charmful baryonic decays}
In this section we shall study the three-body charmful baryonic
$B$ decays: $B^-\to \Lambda_c^+\bar p\pi^-$, $\ov
B^0\to\Sigma_c^{++}\bar p\pi^-$, $\ov B^0\to\Sigma_c^0\bar p\pi^+$
and $B^-\to\Sigma_c^0\bar p\pi^0$.

\subsection{$B^-\to\Lambda_c^+\bar p\pi^-$}
This decay mode has been studied in \cite{CYcharmful} by us.
However, we have missed an important meson-pole contribution
arising from the external $W$-emission diagram. As we shall see
later, this meson-pole effect dominates the decay $\ov
B^0\to\Sigma_c^{++}\bar p\pi^-$.

The decay $B^-\to\Lambda_c^+\bar p\pi^-$ receives resonant and
nonresonant contributions:
 \be
 \Gamma(B^-\to\Lambda_c^+\bar p\pi^-) &=& \Gamma(B^-\to\Lambda_c^+\bar
 p\pi^-)_{\rm nonr} +\Gamma(B^-\to\Sigma_c^0\bar
 p\to\Lambda_c^+\bar p\pi^-) \non \\ &+& \Gamma(B^-\to\Lambda_c^+\bar
 \Delta^{--}\to\Lambda_c^+\bar p\pi^-).
 \en
As the resonant contributions $B^-\to\Sigma_c^0\bar p$ and
$B^-\to\Lambda_c^+\bar\Delta^{--}$ are discussed in the last
section, here we will focus on the nonresonant contribution.

The quark diagrams and the corresponding pole diagrams for $B^-\to
\Lambda_c \,\bar p\,\pi^-$ are shown in Fig. 3. There exist two
distinct internal $W$ emissions and only one of them is
factorizable, namely, Fig. 3(b). The external $W$-emission diagram
Fig. 3(a) is of course factorizable. The factorizable amplitude
reads
 \be \label{factamp}
 A(B^-\to\Lambda_c^+\bar p\pi^-)_{\rm fact} &=&
 {G_F\over\sqrt{2}}V_{cb}V_{ud}^*\Big\{a_1\la \pi^-|(\bar
 du)|0\ra\la \Lambda_c^+\bar p|(\bar cb)|B^-\ra \non\\
 &+& a_2\la\pi^-|(\bar db)|B^-\ra\la \Lambda_c^+\bar p|(\bar
 cu)|0\ra\Big\} \equiv  A_1+A_2,
 \en
where $a_1=c_1^{\rm eff}+c_2^{\rm eff}/N_c$. Let us first consider
the factorizable amplitude $A_2$, as shown in Fig.~3(b), which has
the expression
 \be
 A_2={G_F\over\sqrt{2}}V_{ud}V_{cb}^*\,a_2\bar
 u_{\Lambda_c}\left[(ap\!\!\!/_\pi+b)-(cp\!\!\!/_\pi+d)\gamma_5\right]v_{\bar p},
 \en
where
 \be
 a&=&2f_1^{\Lambda_cp}(t)F_1^{B\pi}(t)+2f_2^{\Lambda_cp}(t)F_1^{B\pi}(t), \non \\
 b&=& (m_{\Lambda_c}-m_p)f_1^{\Lambda_cp}(t)\left[F_1^{B\pi}(t)+(F^{B\pi}_0(t)
 -F^{B\pi}_1(t)){m_B^2-m_\pi^2\over t}\right]  \non \\
 &&-2f_2^{\Lambda_cp}(t)F_1^{B\pi}(t)(p_{\Lambda_c}-p_p)\cdot
 p_\pi/(m_{\Lambda_c}+m_p)
 +f_3^{\Lambda_cp}(t)F_0^{B\pi}(t)(m_B^2-m_\pi^2)/(m_{\Lambda_c}+m_p), \non \\
 c &=& 2g_1^{\Lambda_cp}(t)F_1^{B\pi}(t)+2g_2^{\Lambda_cp}(t)F_1^{B\pi}(t)
(m_{\Lambda_c}-m_p)/(m_{\Lambda_c}+m_p),  \\
 d&=&(m_{\Lambda_c}+m_p)g_1^{\Lambda_cp}(t)\left[F_1^{B\pi}(t)+
 (F^{B\pi}_0(t)-F^{B\pi}_1(t)){m_B^2-m_\pi^2\over
 t}\right]  \non \\
 &&-2g_2^{\Lambda_cp}(t)F_1^{B\pi}(t)(p_{\Lambda_c}-p_p)\cdot
 p_\pi/(m_{\Lambda_c}+m_p)+g_3^{\Lambda_cp}(t)F_0^{B\pi}(t)
 (m_B^2-m_\pi^2)/(m_{\Lambda_c}+m_p), \non
 \en
and $t\equiv q^2=(p_B-p_\pi)^2=(p_{\Lambda_c}+p_{\bar p})^2$, and
we have employed the baryonic form factors defined by
  \be \label{Lamcpfm}
 \la\Lambda_c^+\bar p|(\bar cu)|0\ra &=&
 \bar u_{\Lambda_c}\Bigg\{f_1^{\Lambda_cp}(q^2)\gamma_\mu
 +i{f_2^{\Lambda_cp}(q^2)\over m_{\Lambda_c}+m_p }
 \sigma_{\mu\nu}q^\nu+{f_3^{\Lambda_cp}(q^2)\over m_{\Lambda_c}+m_p}q_{\mu} \non
 \\ &&
 -\Big[g_1^{\Lambda_cp}(q^2)\gamma_\mu+i{g_2^{\Lambda_cp}(q^2)\over
 m_{\Lambda_c}+m_p} \sigma_{\mu\nu}q^\nu+{g_3^{\Lambda_cp}(q^2)
 \over m_{\Lambda_c}+m_p}q_{\mu}\Big]
 \gamma_5\Bigg\}v_{\bar p},
 \en
and the mesonic form factors given by \cite{BSW}
  \be
 \la\pi^-(p_\pi)|(\bar
 db)|B^-(p_B)\ra=F_1^{B\pi}(q^2)(p_B+p_\pi)_\mu+\left(F_0^{B\pi}
 (q^2)-F_1^{B\pi}(q^2)\right){m_B^2-m_\pi^2\over q^2}q_\mu.
 \en

\begin{figure}[t]
\hspace{0cm}\centerline{\psfig{figure=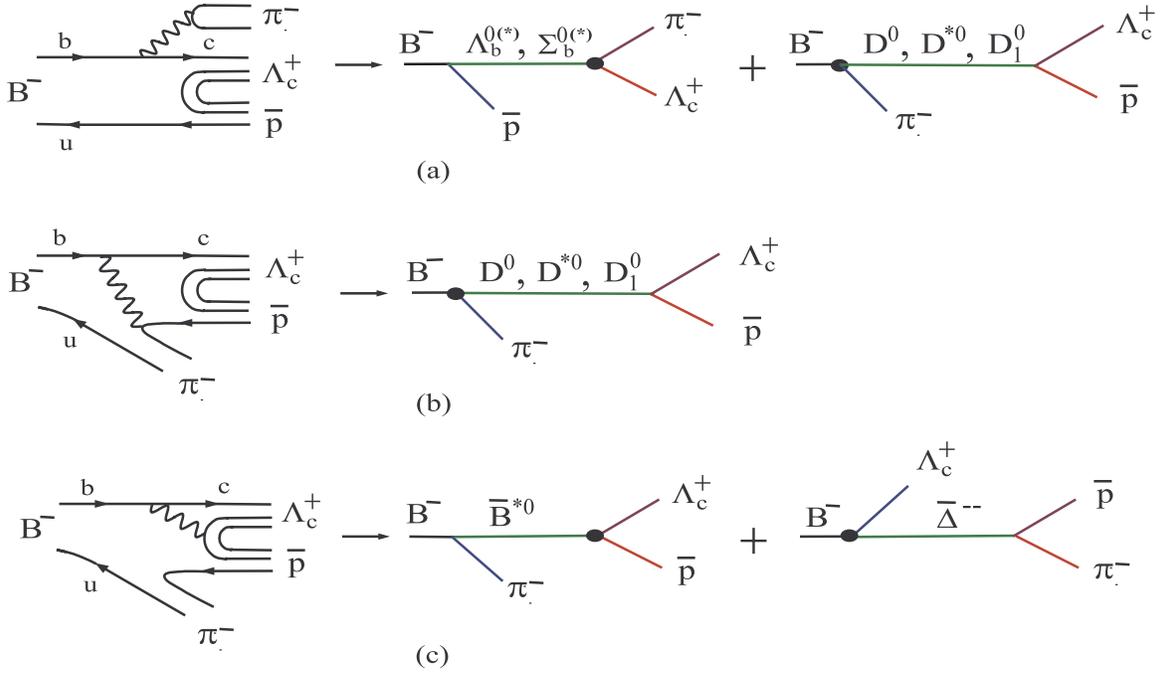,height=3.8in}}
\vspace{0.5cm}
    \caption{{\small Quark and pole diagrams for $B^-\to\Lambda_c^+\bar
    p\pi^-$,
    where the solid blob denotes the weak vertex.
    (a) and (b) correspond to factorizable external and internal $W$-emission
    contributions, respectively, while (c) to nonfactorizable internal $W$-emission
    diagrams. Note that the charmed-meson pole diagram in (a) is color allowed, while it is
    color suppressed in (b).
    }}
\end{figure}

As for the factorizable amplitude $A_1$, since in practice we do
not know how to evaluate the 3-body hadronic matrix element
$\la\Lambda_c^+\bar p|(\bar cb)|B^-\ra$ at the quark level, we
will instead evaluate the corresponding two low-lying pole
diagrams for the external $W$-emission as depicted in Fig. 3(a):
(i) the baryon pole diagram with strong process $B^-\to
\Lambda_b^{0(*)}\bar p$ followed by the weak decay
$\Lambda_b^{0(*)}\to\Lambda_c^+\pi^-$, and (ii) the meson pole
diagram with the color-allowed weak process
$B^-\to\{D^0,D^{*0},D_1^0\}\pi^-$ followed by the strong reaction
$\{D^0,D^{*0},D_1^0\}\to\Lambda_c^+\bar p$. We consider the baryon
pole contribution first. Its amplitude is given by
 \be
 A_{1\B} &=& -{G_F\over\sqrt{2}}V_{ud}V_{cb}^*\,g_{\Lambda_b \to B^-p}f_\pi\,a_1\,\bar
 u_{\Lambda_c}\Big\{f_1^{\Lambda_b\Lambda_c}(m_\pi^2)[2p_\pi\cdot
 p_{\Lambda_c}+p\!\!\!/_\pi(m_{\Lambda_b}-m_{\Lambda_c})]\gamma_5  \non \\
 && +g_1^{\Lambda_b\Lambda_c}(m_\pi^2)[2p_\pi\cdot
 p_{\Lambda_c}-p\!\!\!/_\pi(m_{\Lambda_b}+m_{\Lambda_c})]\Big\}
 v_{\bar p}\times{1\over
 (p_{\Lambda_c}+p_\pi)^2-m_{\Lambda_b}^2 },
 \en
where we have applied factorization to the weak decay
$\Lambda_b^0\to\Lambda_c^+\pi^-$. Note that the intermediate
states $\Sigma_b^0$ and $\Sigma_b^{0*}$ also do not contribute to
$A_1$ under the factorization approximation because the weak
transition  $\la \Lambda_c|(\bar cb)|\Sigma_b^{0(*)}\ra$ is
prohibited as $\Sigma_b$ and $\Sigma_b^{*}$ are sextet bottom
baryons whereas $\Lambda_c$ is an anti-triplet charmed baryon.

The meson pole contribution from Fig. 3(a) consists of the
pseudoscalar meson $D^0$, the vector meson $D^{*0}$ and the
axial-vector meson $D_1^0(2400)$. Note that the weak decay process
$B^-\to\{D^0,D^{*0},D_1^0\}\pi^-$ in Fig. 3(a) is color allowed,
namely, its amplitude is proportional to $a_1$, while the same
process in Fig. 3(b), being proportional to $a_2$, is color
suppressed.   The charmed-meson pole amplitude has the form
 \be
 A_{1\cal M} &=&
 {G_F\over\sqrt{2}}V_{ud}^*V_{cb}\,a_1\,\la \pi^-|(\bar
 du)|0\ra\Bigg\{ \Big[ \la D^0|(\bar cb)|B^-\ra{i\over
 q^2-m_D^2}g^{\Lambda_c\to pD^0}\,
 \bar u_{\Lambda_c}\gamma_5 v_{\bar p} \non \\
 &+& \la D^{*0}|(\bar cb)|B^-\ra
 {i\over q^2-m_{D^*}^2}\,\bar u_{\Lambda_c}\,i\,\vp^{\nu}_{D^*}\left(g_1^{\Lambda_c^{+}
 \to pD^{*0}}\gamma_\nu+i{g_2^{\Lambda_c^{+}\to pD^{*0}}\over
 m_{\Lambda_c}+m_p}\sigma_{\nu\lambda}q^\lambda\right)v_{\bar p} \non \\
 &+& \la D_1^{0}|(\bar cb)|B^-\ra
 {i\over q^2-m_{D_1}^2}\,\bar u_{\Lambda_c}\,i\,\vp^{\nu}_{D_1}\left(h_1^{\Lambda_c^{+}
 \to pD_1^{0}}\gamma_\nu+i{h_2^{\Lambda_c^{+}\to pD_1^{0}}\over
 m_{\Lambda_c}+m_p}\sigma_{\nu\lambda}q^\lambda\right)\gamma_5 v_{\bar p}
  \Bigg\},
 \en
where $q=p_B-p_\pi=p_{\Lambda_c}+p_{\bar p}$, and $g$,
$g_1,\,g_2$, $h_1,\,h_2$ are the unknown strong couplings. After
some manipulation we obtain
 \be
  A_{1\cal M} &=&
 -{G_F\over\sqrt{2}}V_{ud}^*V_{cb}\,f_\pi a_1\Bigg\{
 (m_B^2-m_D^2)F_0^{BD}(m_\pi^2){g^{\Lambda_c\to pD^0}\over q^2-m_D^2}
 \,\bar u_{\Lambda_c}\gamma_5 v_{\bar p}   \non \\
 &+& {2 m_{D^*}\over q^2-m_{D^*}^2} A_0^{BD^*}(m_\pi^2)p^\mu_B\left(-g_{\mu\nu}
 +{q_\mu q_\nu\over m_{D^*}^2}\right)
 \bar u_{\Lambda_c}\left(g_1^{\Lambda_c^+
 \to pD^{*0}}\gamma^\nu+i{g_2^{\Lambda_c^{+}\to pD^{*0}}\over
 m_{\Lambda_c}+m_p}\sigma^{\nu\lambda}q_\lambda\right)v_{\bar p}
 \non \\ &+& {2 m_{D_1}\over q^2-m_{D_1}^2} V_0^{BD_1}(m_\pi^2)p^\mu_B\left(-g_{\mu\nu}
 +{q_\mu q_\nu\over m_{D_1}^2}\right)
 \bar u_{\Lambda_c}\left(h_1^{\Lambda_c^{+}
 \to pD^0_1}\gamma^\nu+i{h_2^{\Lambda_c^{+}\to pD_1^0}\over
 m_{\Lambda_c}+m_p}\sigma^{\nu\lambda}q_\lambda\right)\gamma_5 v_{\bar p}
 \Bigg\}, \non \\
 \en
where we have employed the form factors defined by\footnote{Our
definition for $B\to D^*$ form factors is the same as \cite{BSW}
except for a sign difference for the matrix elements of the
axial-vector current. This sign change is required in order to
ensure positive form factors as one can check via heavy quark
symmetry or the QCD sum rule analysis.}
 \be
 \la D^{*0}(p_{D^*},\vp)|(\bar cb)_\vma|B^-(p_B)\ra &=& {2\over
 m_B+m_{D^*}}\epsilon_{\mu\nu\alpha\beta}
\vp^{*\nu}p^\alpha_{D^*} p^\beta_B V^{BD^*}(q^2)  \non \\ &-&i
\Bigg\{(m_B+m_{D^*}) \vp^*_\mu A_1^{BD^*}(q^2)  - {\vp^*\cdot
p_B\over m_B+m_{D^*}}(p_B+p_{D^*})_\mu A_2^{BD^*}(q^2) \non \\
&-& 2m_{D_1} {\vp^*\cdot p_B\over
q^2}q_\mu\left[A_3^{BD^*}(q^2)-A_0^{BD^*}(q^2)\right]\Bigg\},
 \en
with
  \be
A_3^{BD^*}(q^2)=\,{m_B+m_{D^*}\over
2m_{D^*}}\,A^{BD^*}_1(q^2)-{m_B-m_{D^*}\over
2m_{D^*}}\,A^{BD^*}_2(q^2),
 \en
and
  \be \label{formBa1}
 \la D_1^0(p_{D_1},\vp)|(\bar cb)_\vma|B^-(p_B)\ra &=& {2\over
 m_B+m_{D_1}}\epsilon_{\mu\nu\alpha\beta}
\vp^{*\nu}p^\alpha_{D_1} p^\beta_B A^{BD_1}(q^2)  \non \\ &-&i
\Bigg\{(m_B+m_{D_1}) \vp^*_\mu V_1^{BD_1}(q^2)  - {\vp^*\cdot
p_B\over m_B+m_{D_1}}(p_B+p_{D_1})_\mu V_2^{BD_1}(q^2) \non \\
&-& 2m_{D_1} {\vp^*\cdot p_B\over
q^2}q_\mu\left[V_3^{BD_1}(q^2)-V_0^{BD_1}(q^2)\right]\Bigg\},
 \en
with
  \be
V_3^{BD_1}(q^2)=\,{m_B+m_{D_1}\over
2m_V}\,V^{BD_1}_1(q^2)-{m_B-m_{D_1}\over
2m_{D_1}}\,V^{BD_1}_2(q^2)
 \en
and $A^{BD^*}_3(0)=A^{BD^*}_0(0)$ as well as
$V_3^{BD_1}(0)=V^{BD_1}_0(0)$. The above amplitude can be further
simplified by applying the Gordon decomposition as
  \be
  A_{1\cal M} &=&
 {G_F\over\sqrt{2}}V_{ud}^*V_{cb}\,f_\pi a_1\Bigg\{
 -(m_B^2-m_D^2)F_0^{BD}(m_\pi^2){g^{\Lambda_c\to pD^0}\over q^2-m_D^2}
 \,\bar u_{\Lambda_c}\gamma_5 v_{\bar p}   \non \\
 &+& {2 m_{D^*}\over q^2-m_{D^*}^2} A_0^{BD^*}(m_\pi^2)
 \bar u_{\Lambda_c}\Big[ (g_1^{\Lambda_c\to pD^{*0}}+g_2^{\Lambda_c\to pD^{*0}})
 p\!\!\!/_B \non \\ &-& g_1^{\Lambda_c\to pD^{*0}}{(p_B\cdot q)(m_{\Lambda_c}-m_p)
 \over m^2_{D^*}}-g_2^{\Lambda_c\to pD^{*0}}{p_B\cdot(p_{\Lambda_c}-p_{\bar p})
 \over m_{\Lambda_c}+m_p}\Big]v_{\bar p}
 \non \\ &+& {2 m_{D_1}\over q^2-m_{D_1}^2} V_0^{BD_1}(m_\pi^2)
 \bar u_{\Lambda_c}\Big[ (h_1^{\Lambda_c\to pD_1^0}+{m_{\Lambda_c}-m_p\over
 m_{\Lambda_c}+m_p}\,h_2^{\Lambda_c\to pD_1^0})
 p\!\!\!/_B \non \\ &-& h_1^{\Lambda_c\to pD^0_1}\,{(p_B\cdot q)(m_{\Lambda_c}+m_p)
 \over m^2_{D_1}}-h_2^{\Lambda_c\to pD^0_1}\,{p_B\cdot(p_{\Lambda_c}-p_{\bar p})
 \over m_{\Lambda_c}+m_p}\Big]\gamma_5 v_{\bar p}
 \Bigg\}.
 \en

In order to compute the nonresonant decay rate for
$B^-\to\Lambda_c^+\bar p\pi^-$ we  need to know the strong
couplings $g,\,g_1,\,g_2$, $h_1,h_2$ and their $q^2$ dependence.
Fortunately, this can be achieved by considering the meson-pole
contributions to the factorizable internal $W$-emission as
depicted in Fig. 3(b). In the pole model description, the relevant
intermediate states are $D^0,D^{*0}$ and $D_1^0(2400)$ as shown in
the same figure. The matrix element $\la \Lambda_c^+\bar
p|(V-A)_\mu|0\ra$ then reads
 \be
 \la \Lambda_c^+\bar p|(V-A)_\mu |0\ra_{\rm pole} &=& \bar u_{\Lambda_c}\Bigg\{
 {f_{D^*}m_{D^*}\over q^2-m_{D^*}^2}\left[g_1^{\Lambda_c\to pD^*}
 \gamma_\mu+i{g_2^{\Lambda_c\to pD^*}\over
 m_{\Lambda_c}+m_p}\sigma_{\mu\nu}q^\nu\right]   \non \\
 &-& {f_{D_1} m_{D_1}\over q^2-m_{D_1}^2}\left[h_1^{\Lambda_c\to
 pD_1}\gamma_\mu+i{h_2^{\Lambda_c\to pD^*}\over
 m_{\Lambda_c}+m_p}\sigma_{\mu\nu}q^\nu\right]\gamma_5  \\
 &-& \left[{f_D g^{\Lambda_c\to pD}\over
 q^2-m_D^2} -{f_{D_1}g_1^{\Lambda_c\to pD_1} \over
 q^2-m_{D_1}^2}{m_{\Lambda_c}+m_p\over m_{D_1}}\right] q_\mu\gamma_5\Bigg\} v_{\bar
 p}. \non
 \en
where the decay constants are defined by
 \be
 && \la D(q)|A_\mu|0\ra= -if_Dq_\mu, \quad \la
 D^*(q,\vp)|V_\mu|0\ra= f_{D^*}m_{D^*}\vp_\mu^*, \non \\
 && \la D_1(q,\vp)|A_\mu|0\ra= f_{D_1}m_{D_1}\vp_\mu^*.
 \en
Comparing this with Eq. (\ref{Lamcpfm}) we see that the $D^*$
meson is responsible for the strong couplings $g_1$ and $g_2$,
$D_1(2400)$ for $h_1$ and $h_2$, and $D$ for the coupling $g$.
More precisely,
 \be
 g_1^{\Lambda_c\to pD^*}(q^2) &=& {q^2-m_{D^*}^2\over f_{D^*}
 m_{D^*}}\,f_1^{\Lambda_c p}(q^2), \qquad\quad  g_2^{\Lambda_c\to pD^*}(q^2)
  = {q^2-m_{D^*}^2\over f_{D^*}
 m_{D^*}}\,f_2^{\Lambda_c p}(q^2),  \non \\
 h_1^{\Lambda_c\to pD_1}(q^2) &=& {q^2-m_{D_1}^2\over f_{D_1}
 m_{D_1}}\,g_1^{\Lambda_c p}(q^2), \qquad\quad h_2^{\Lambda_c\to pD_1}(q^2)
 = {q^2-m_{D_1}^2\over f_{D_1}
 m_{D_1}}\,g_2^{\Lambda_c p}(q^2), \non \\
 g^{\Lambda_c\to pD}(q^2)
 &=& {q^2-m_D^2\over
 f_D(m_{\Lambda_c}+m_p)}\,g_3^{\Lambda_c p}(q^2),
 \en
where the $D_1$ pole contribution to $g_3^{\Lambda_cp}$ can be
neglected at the $q^2$ range of interest.

The form factors $f_i$ and $g_i$ for the heavy-to-heavy and
heavy-to-light baryonic transitions at zero recoil have been
computed using the non-relativistic quark model \cite{CT96}. In
principle, HQET puts some constraints on these form factors.
However, it is clear that HQET is not adequate for our purposes:
the predictive power of HQET for the baryon form factors at order
$1/m_Q$ is limited only to the antitriplet-to-antitriplet heavy
baryonic transition. Hence, we will follow \cite{CT96} to apply
the nonrelativistic quark model to evaluate the weak
current-induced baryon form factors at zero recoil in the rest
frame of the heavy parent baryon, where the quark model is most
trustworthy. This quark model approach has the merit that it is
applicable to heavy-to-heavy and heavy-to-light baryonic
transitions at maximum $q^2$. It has been shown in \cite{CT96}
that the quark model predictions agree with HQET for the
antitriplet-to-antitriplet (e.g. $\Lambda_b\to\Lambda_c,~
\Xi_b\to\Xi_c$) form factors to order $1/m_Q$. For sextet
$\Sigma_b\to \Sigma_c$ and $\Omega_b\to\Omega_c$ transitions, the
quark-model results  are also in accord with the HQET predictions
(for details see \cite{Cheng97}). Numerically we have
\cite{Cheng97}
 \be
&&
f_1^{\Lambda_b\Lambda_c}(q^2_m)=g_1^{\Lambda_b\Lambda_c}(q^2_m)=1.02,
 \quad
 f_2^{\Lambda_b\Lambda_c}(q^2_m)=g_3^{\Lambda_b\Lambda_c}(q^2_m)=-0.23,
 \non \\
&&
f_3^{\Lambda_b\Lambda_c}(q^2_m)=g_2^{\Lambda_b\Lambda_c}(q^2_m)=-0.03,
 \en
for the $\Lambda_b\to\Lambda_c$ transition at zero recoil
$q_m^2=(m_{\Lambda_b}-m_{\Lambda_c})^2$, and
\cite{CT96}\footnote{The $\Lambda_c\to p$ form factors
$f_2,f_3,g_2,g_3$ given in Eq. (\ref{fLamcp}) are different from
that in \cite{CYcharmful} owing to a different definition of these
form factors.}
 \be \label{fLamcp}
&& f_1^{\Lambda_cp}(q^2_m)=g_1^{\Lambda_cp}(q^2_m)=0.79,
 \quad
 f_2^{\Lambda_cp}(q^2_m)=g_3^{\Lambda_cp}(q^2_m)=-0.69,
 \non \\
 && f_3^{\Lambda_cp}(q^2_m)=g_2^{\Lambda_cp}(q^2_m)=-0.20,
 \en
for the $\Lambda_c\to p$ transition at
$q^2_m=(m_{\Lambda_c}-m_p)^2$.

Since the calculation for the $q^2$ dependence of form factors is
beyond the scope of the non-relativistic quark model, we will
follow the conventional practice to assume a pole dominance for
the form-factor $q^2$ behavior:
 \be
 f(q^2)=f(q^2_m)\left({1-q^2_m/m^2_V\over 1-q^2/m_V^2} \right)^n\,,\qquad
 g(q^2)=g(q^2_m)
\left({1-q^2_m/m^2_A\over 1-q^2/m_A^2} \right)^n\,,
 \en
where $m_V$ ($m_A$) is the pole mass of the vector (axial-vector)
meson with the same quantum number as the current under
consideration. The function
 \be
 G(q^2)=\left({1-q^2_m/m^2_{\rm
pole}\over 1-q^2/m_{\rm pole}^2} \right)^n
 \en
plays the role of the baryon Isgur-Wise function $\zeta(\omega)$
for the $\Lambda_Q\to \Lambda_{Q'}$ transition, namely, $G=1$ at
$q^2=q^2_m$. However, whether the $q^2$ dependence is monopole
($n=1$) or dipole ($n=2$) for heavy-to-heavy transitions is not
clear. Hence we shall use both monopole and dipole dependence in
ensuing calculations. Moreover, one should bear in mind that the
$q^2$ behavior of form factors is probably more complicated and it
is likely that a simple pole dominance only applies to a certain
$q^2$ region, especially for the heavy-to-light transition. We
will use the pole masses $m_V=2.01$ GeV and $m_A=2.42$ GeV for the
$\Lambda_c\to p$ transition and $m_V=6.34$ GeV, $m_A=6.73$ GeV for
$\Lambda_b\to\Lambda_c$ and $\Sigma_b\to\Sigma_c$ transitions.

For the form factors $F_{0,1}^{B\pi}(q^2)$ we consider the
Melikhov-Stech (MS) model based on the constituent quark picture
\cite{MS}. Although the form factor $q^2$ dependence is in general
model dependent, it should be stressed that $F_1^{B\pi}(q^2)$
increases with $q^2$ more rapidly than $F_0^{B\pi}(q^2)$ as
required by heavy quark symmetry.

The total decay rate for the process $B^-(p_B)\to
\Lambda_c(p_1)+\bar p(p_2)+\pi^-(p_3)$ is computed by
 \be \Gamma = {1\over (2\pi)^3}\,{1\over
32m_B^3}\int |A|^2dm_{12}^2dm_{23}^2,
 \en
where $m_{ij}^2=(p_i+p_j)^2$ with $p_3=p_\pi$.  Under naive
factorization, the parameter $a_2$ appearing in Eq.
(\ref{factamp}) is numerically equal to 0.024, which is very small
compared to the value of $a_2=0.40-0.55$ extracted from $\ov B^0
\to D^{0(*)}\pi^0$ decays \cite{Cheng01} and $|a_2|=0.26\pm 0.02$
from the $B\to\J K$ decay \cite{a1a2}. Since $a_2$ may receive
sizable contributions from the pole diagram Fig. 3(c), we will
thus treat $a_2$ as a free parameter and take $a_2=0.30$ as an
illustration.

Collecting everything together we obtain numerically\footnote{If
we use $|g_{\Lambda_b\to B^-p}|=16$ as in \cite{CYcharmful}, then
we will have $\B(B^-\to\Lambda_c^+\bar p\pi^-)_{\rm
nonr}=5.0\times 10^{-4}$ for $n=1$ and $8.3\times 10^{-4}$ for
$n=2$.}
 \be \label{Lamcppi}
 \B(B^-\to\Lambda_c^+\bar p\pi^-)_{\rm nonr}=\cases{ 1.8\times 10^{-4}
 & for~$n=1$, \cr 2.6\times 10^{-4} & for~$n=2$, }
 \en
where we have used $g_{\Lambda_b\to B^-p}=-7$ (or
$g_{\Sigma_b^+\to\ov B^0p}=5$), $f_D=200$ MeV,
$f_{D^*}=f_{D_1}=230$ MeV and $V_0^{BD_1}(0)=0.37$. It should be
stressed that the sign of the strong coupling $g_{\Lambda_b\to
B^-p}$ must be negative and hence $g_{\Sigma_b^+\to\ov B^0p}$ has
to be positive [see Eq. (\ref{3P0m})] so that the interference
between meson and baryon pole contributions is destructive for
$n=1$ and constructive for $n=2$. Indeed, if $g_{\Lambda_b\to
B^-p}=7$ is employed, one will have a branching ratio of order
$9.6\times 10^{-4}$ for $n=1$ and $4.7\times 10^{-5}$ for $n=2$,
in disagreement with experiment. We shall see below that the
prediction of the $\Sigma_c^{++}\bar p\pi^-$ rate is consistent
with experiment only for $n=1$.  Adding the resonant contributions
from $\Sigma_c^0\bar p$ and $\Lambda_c^+\bar\Delta^{--}$ as
discussed in Sec. II, we are led to
 \be
 \B(B^-\to\Lambda_c^+\bar p\pi^-)\approx 2.4\times 10^{-4}.
 \en
Note that the resonant contributions account for about one quarter
of the total decay rate.

In \cite{CYcharmful} we obtained a branching ratio of order
$(4.9\sim 9.2)\times 10^{-4}$ for $|g_{\Lambda_b\to B^-p}|=16$.
Our present results (\ref{Lamcppi}) are smaller for two reasons:
(i) The large strong coupling  $|g_{\Lambda_b\to B^-p}|=16$ will
lead to a too large $B^-\to\Sigma_c^0\bar p$ which is ruled out by
experiment. Therefore we use $|g_{\Lambda_b\to B^-p}|=7$, obtained
by fitting to the observed central value of $B^-\to\Sigma_c^0\bar
p$. The branching ratio due to the baryon poles becomes $1.7\times
10^{-4}$ for $n=1$ and $1.1\times 10^{-4}$ for $n=2$. (ii) The
color-allowed charmed-meson pole contribution to the branching
ratio is $3.6\times 10^{-4}$ for $n=1$ and $0.5\times 10^{-4}$ for
$n=2$. It has a destructive (constructive) interference with
baryon pole contributions for $n=1$ ($n=2$).
%Note that the
%destructive interference between baryon and meson pole
%contributions for $n=1$ does not lead to a much smaller branching
%ratio because the baryon pole dominates the vector part, while
%baryon and meson poles contribute in a similar rate to the
%axial-vector part for $n=1$.

\subsection{$\ov B^0\to\Sigma_c^{++}\bar p\pi^-$}
The three-body mode $\ov B^0\to \Lambda_c^{++} \,\bar p\,\pi^-$
does receive factorizable internal $W$-emission and $W$-exchange
contributions:
 \be
 A(\ov B^0\to\Sigma_c^{++}\bar p\pi^-)_{\rm fact} &=&
 {G_F\over\sqrt{2}}V_{cb}V_{ud}^*\,\Big\{a_1\la \pi^-|(\bar
 du)|0\ra\la \Sigma_c^{++}\bar p|(\bar cb)|\ov B^0\ra \non \\
 &+& a_2\la\Sigma_c^{++}\bar p\pi^-|(\bar cu)|0\ra\la 0|(\bar
 db)|\ov B^0\ra\Big\}.
 \en
As before, we do not know how to evaluate the 3-body hadronic
matrix element $\la\Sigma_c^{++}\bar p|(\bar cb)|\ov B^0\ra$ at
the quark level. Thus we will instead evaluate the corresponding
two low-lying pole diagrams for the external $W$-emission: (i) the
baryon pole diagram with strong process $\ov B^0\to
\Sigma_b^{(*)+}\bar p$ followed by the weak decay
$\Sigma_b^{(*)+}\to\Sigma_c^{++}\pi^-$ [see Fig. 4(a)], and (ii)
the meson pole diagram with the weak process $\ov B^0\to
\{D^+,D^{*+},D_1^+\}\pi^-$ followed by the strong reaction
$\{D^+,D^{*+},D^+_1\}\to\Sigma_c^{++}\bar p$ [Fig. 4(a)].  We
consider the baryon pole contribution first. Its amplitude is
given by
 \be
 A(\ov B^0\to \Sigma_c^{++}\bar p\pi^-)_{\cal B}  &=&
 -{G_F\over\sqrt{2}}V_{ud}V_{cb}^*\,g_{\Sigma_b^+\to \ov B^0p}f_\pi\,a_1\,\bar
 u_{\Sigma_c}\Big\{f_1^{\Sigma_b\Sigma_c}(m_\pi^2)[2p_\pi\cdot
 p_{\Sigma_c}+p\!\!\!/_\pi(m_{\Sigma_b}-m_{\Sigma_c})]\gamma_5  \non \\
 && +g_1^{\Sigma_b\Sigma_c}(m_\pi^2)[2p_\pi\cdot
 p_{\Sigma_c}-p\!\!\!/_\pi(m_{\Sigma_b}+m_{\Sigma_c})]\Big\}
 v_{\bar p}\times{1\over
 (p_{\Sigma_c}+p_\pi)^2-m_{\Sigma_b}^2 },
 \en
where we have applied factorization to the weak decay
$\Sigma_b^+\to\Sigma_c^{++}\pi^-$.

\begin{figure}[t]
\hspace{0.5cm}\psfig{figure=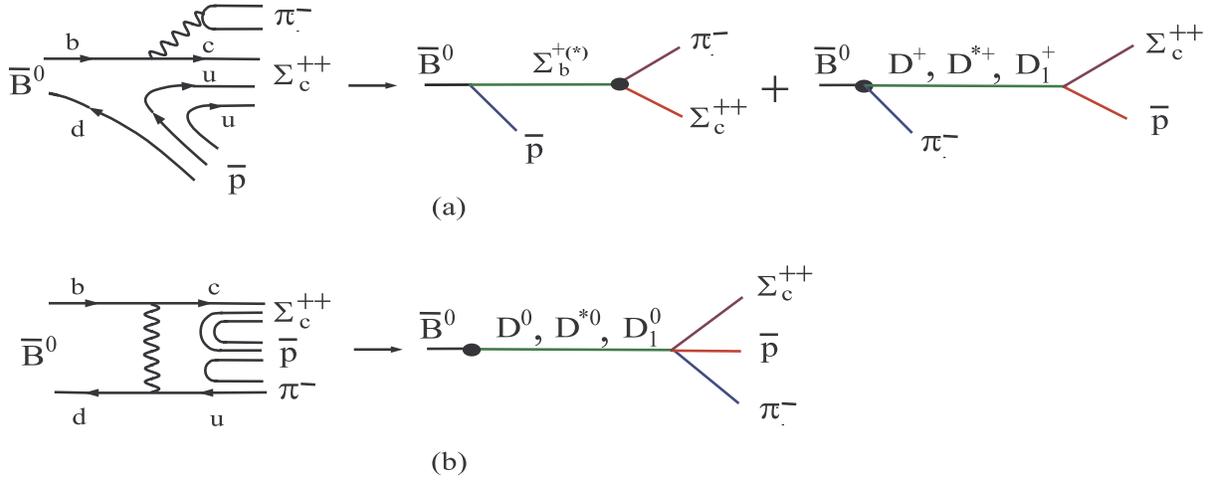,height=2.8in}
\vspace{0.5cm}
    \caption{{\small Quark and pole diagrams for $\ov B^0\to\Sigma_c^{++}\bar p\pi^-$
    where the solid blob denotes the weak vertex.
    (a) and (b) correspond to the external $W$-emission and
    $W$-exchange contributions, respectively.
    }}
\end{figure}

The heavy-to-heavy transition $\Sigma_b\to\Sigma_c$ at zero recoil
is predicted by HQET to be (see  e.g. \cite{Neubert})
 \be
f_1^{\Sigma_b\Sigma_c}(q_m^2) &=& -{1\over
3}\left[1-(m_{\Sigma_b}+m_{\Sigma
_c})\left({1\over m_{\Sigma_b}}+{1\over m_{\Sigma_c}}\right)\right],  \non \\
f_2^{\Sigma_b\Sigma_c}(q_m^2) &=& {1\over 3}\left({1\over
m_{\Sigma_b}}+
{1\over m_{\Sigma_c}}\right)(m_{\Sigma_b}+m_{\Sigma_c}),  \non \\
f_3^{\Sigma_b\Sigma_c}(q_m^2) &=& {1\over 3}\left({1\over
m_{\Sigma_b}}-{1\over m_{\Sigma_c}}\right)(m_{\Sigma_b}+m_{\Sigma_c}),   \\
g_1^{\Sigma_b\Sigma_c}(q^2_m) &=& -{1\over
3},~~~~g_2^{\Sigma_b\Sigma_c}
(q^2_m)=g_3^{\Sigma_b\Sigma_c}(q^2_m)=0,  \non
 \en
where $q_m^2=(m_{\Sigma_b}-m_{\Sigma_c})^2$. Numerically we find a
small branching ratio
 \be
 \B(\ov B^0\to\Sigma_c^{++}\bar p\pi^-)_\B=\cases{0.45\times
 10^{-4}& $n=1$, \cr 0.24\times 10^{-4} & $n=2$ }
 \en
arising from the baryon poles. Comparing to the experimental value
(see Table I), it is obvious that the baryon pole contribution
alone is not adequate to account for the data and it is necessary
to take into account the meson pole contribution.

The meson pole contribution from Fig. 4(a) is
 \be
 A(\ov B^0\to \Sigma_c^{++}\bar p\pi^-)_{\cal M} &=&
 {G_F\over\sqrt{2}}V_{ud}^*V_{cb}\,a_1\,\la \pi^-|(\bar
 du)|0\ra\Bigg\{ \Big[ \la D^+|(\bar cb)|\ov B^0\ra{i\over
 q^2-m_D^2}g^{\Sigma_c^{++}\to pD^+}
 \bar u_{\Sigma_c}\gamma_5 v_{\bar p} \non \\
 &+& \la D^{*+}|(\bar cb)|\ov B^0\ra
 {i\over q^2-m_{D^*}^2}\,\bar u_{\Sigma_c}\,i\,\vp^{\nu}_{D^*}\left(g_1^{\Sigma_c^{++}
 \to pD^{*+}}\gamma_\nu+i{g_2^{\Sigma_c^{++}\to pD^{*+}}\over
 m_{\Sigma_c}+m_p}\sigma_{\nu\lambda}q^\lambda\right)v_{\bar p} \non \\
 &+& \la D^+_1|(\bar cb)|\ov B^0\ra
 {i\over q^2-m_{D_1}^2}\,\bar u_{\Sigma_c}\,i\,\vp^{\nu}_{D_1}\left(h_1^{\Sigma_c^{++}
 \to pD_1^+}\gamma_\nu+i{h_2^{\Sigma_c^{++}\to pD_1^+}\over
 m_{\Sigma_c}+m_p}\sigma_{\nu\lambda}q^\lambda\right)v_{\bar p}
  \Bigg\}. \non \\
 \en
After some manipulation we obtain
 \be
  A(\ov B^0\to \Sigma_c^{++}\bar p\pi^-)_{\cal M} &=&
  {G_F\over\sqrt{2}}V_{ud}^*V_{cb}\,f_\pi a_1\Bigg\{-
 (m_B^2-m_D^2)F_0^{BD}(m_\pi^2){g^{\Sigma_c^{++}\to pD^+}\over q^2-m_D^2}
 \,\bar u_{\Sigma_c}\gamma_5 v_{\bar p}   \non \\
 &+& {2 m_{D^*}\over q^2-m_{D^*}^2} A_0^{BD^*}(m_\pi^2)
 \bar u_{\Sigma_c}\Big[ (g_1^{\Sigma_c^{++}\to pD^{*+}}+g_2^{\Sigma_c^{++}\to pD^{*+}})
 p\!\!\!/_B \non \\ &-& g_1^{\Sigma_c^{++}\to pD^{*+}}{(p_B\cdot q)(m_{\Sigma_c}-m_p)
 \over m^2_{D^*}}-g_2^{\Sigma_c^{++}\to pD^{*+}}{p_B\cdot(p_{\Sigma_c}-p_{\bar p})
 \over m_{\Sigma_c}+m_p}\Big]v_{\bar p}
 \non \\ &+& {2 m_{D_1}\over q^2-m_{D_1}^2} V_0^{BD_1}(m_\pi^2)
 \bar u_{\Sigma_c}\Big[ (h_1^{\Sigma_c^{++}\to pD_1^+}
 +{m_{\Sigma_c}-m_p\over m_{\Sigma_c}+m_p}\,h_2^{\Sigma_c^{++}\to pD_1^+})
 p\!\!\!/_B \non \\ &-& h_1^{\Sigma_c^{++}\to pD_1^+}\,{(p_B\cdot q)(m_{\Sigma_c}+m_p)
 \over m^2_{D_1}}-h_2^{\Sigma_c^{++}\to pD_1^+}\,{p_B\cdot(p_{\Sigma_c}-p_{\bar p})
 \over m_{\Sigma_c}+m_p}\Big]\gamma_5 v_{\bar p}
 \Bigg\}. \non \\
 \en

There exist five unknown strong couplings $g^{\Sigma_c^{++}\to
pD^+}$, $g_{_{1,2}}^{\Sigma_c^{++}\to pD^{*+}}$ and
$h_{_{1,2}}^{\Sigma_c^{++}\to pD^{+}_1}$ which are $q^2$
dependent. To determine these couplings we apply the $^3P_0$
quark-pair-creation model to obtain
 \be \label{SigmacpD}
 g_{_{1,2}}^{\Sigma_c^{++}\to pD^*}(q^2) &=& \sqrt{3\over 2}\,g_{_{1,2}}^{\Lambda_c^+\to
 pD^*}(q^2), \qquad\quad
 h_{_{1,2}}^{\Sigma_c^{++}\to pD_1}(q^2) = \sqrt{3\over 2}\,h_{_{1,2}}^{\Lambda_c^+\to
 pD_1}(q^2), \non \\
 g^{\Sigma_c^{++}\to pD}(q^2) &=& -3\sqrt{3\over 2}\,g^{\Lambda_c^+\to
 pD}(q^2).
 \en
As noted in passing, the $^3P_0$ model is perhaps reliable only in
the low energy regime. Nevertheless, we will use Eq.
(\ref{SigmacpD}) for an estimation. We obtain numerically
 \be
 \B(\ov B^0\to\Sigma_c^{++}\bar p\pi^-)=\cases{ 4.5\times 10^{-4}
 & for~$n=1$, \cr 4.3\times 10^{-6} & for~$n=2$, }
 \en
for $g_{\Sigma_b^+\to\ov B^0p}=5$. Note that the interference
between meson and baryon pole contributions is constructive
(destructive) for $n=1$ ($n=2$), opposite to the case of
$\Lambda_c^+\bar p\pi^-$. Evidently, $n=1$ is favored by the
measurements of Belle and CLEO. Therefore, we conclude that $\ov
B^0\to\Sigma_c^{++}\bar p\pi^-$ has a rate slightly larger than
$B^-\to\Lambda_c^+\bar p\pi^-$.

\subsection{$\ov B^0\to\Sigma_c^0\bar p\pi^+$ and $B^-\to\Sigma_c^0\bar p\pi^0$}
The decays $\ov B^0\to\Sigma_c^0\bar p\pi^+$ and
$B^-\to\Sigma_c^0\bar p\pi^0$ proceed via the nonfactorizable
internal $W$-emission (see Figs. 5 and 6). Naively one may argue
that they are color suppressed relative to $\ov
B^0\to\Sigma_c^{++}\bar p\pi^-$. However, it may not be the case
for baryonic $B$ decays. To demonstrate this, let us take a look
at Fig. 3(b) which proceeds via an internal $W$-emission. This
diagram is color suppressed because in order to form the $\pi^-$
and $\Lambda_c^+\bar p$, the color of the spectator $\bar u$ quark
has to be matched with that of the $d$ quark created from the $b$
quark decay, and similarly the color of the $c$ quark has to be
matched with that of the $\bar u$ quark created from the $b$ quark
decay. In the effective Hamiltonian approach, the factorizable
amplitude is proportional to $a_2\la \pi^-|(\bar
db)|B^-\ra\la\Lambda_c^+\bar p|(\bar cu)|0\ra$, where the
coefficient $a_2$ is equivalent to 1/3 in the absence of strong
interactions. On the contrary, the other internal $W$-emission
diagram Fig. 3(c) is not color suppressed because the color wave
function of the baryon is totally antisymmetric and hence the
color of the $c$ quark must be different from that of the $d$
quark created from the $b$ quark decay. Likewise, Figs. 5 and 6
are not color suppressed. Indeed, as shown in Sec. II, the weak
baryon-baryon transition is found to be proportional to $c_1-c_2$
rather than to $c_2+c_1/N_c$.

\begin{figure}[t]
\hspace{0cm}\centerline{\psfig{figure=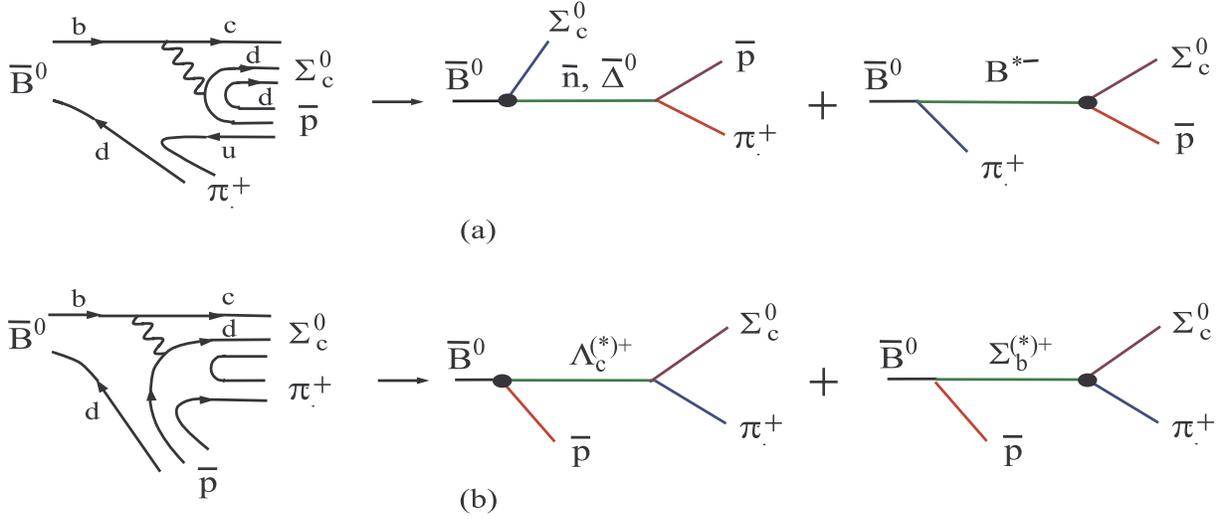,height=3.0in}}
\vspace{0.5cm}
    \caption{{\small Quark and pole diagrams for $\ov B^0\to\Sigma_c^0\bar
    p\pi^+$, where the solid blob denotes the weak vertex.
    }}
\end{figure}

Theoretically, it is not easy to estimate the pole contributions
as the weak decay processes, for example $\ov B^0\to\Sigma_c^0\bar
n$ and $\Sigma_b^+\to\Sigma_c^0\pi^+$ in Fig. 5, are not
factorizable. This renders the calculation difficult. Among the
pole diagrams, only the intermediate $\bar n$ state in Fig. 5(a)
and the $\bar p$ state in Fig. 6 can be reliably estimated since
the involved weak transitions are already discussed in Sec. II and
moreover the strong $\pi NN$ coupling can be related to the
nucleon-nucleon form factor $g_3^{np}(q^2)$ and hence its $q^2$
dependence can be determined \cite{CYBDNN}.

Consider the decay $\ov B^0\to\Sigma_c^0\bar p\pi^+$ first. The
decay amplitude of the $\bar n$ pole diagram reads
 \be
 A(\ov B^0\to\Sigma_c^0\bar p\pi^+)_{{\bar n}-{\rm pole}} &=&
 \sqrt{2}\,g_{\pi NN}(q^2)B(\ov B^0\to\Sigma_c^0\bar n)\bar
 u_{\Sigma_c}p\!\!\!/_\pi v_{\bar p}\times
 {1\over q^2-m_N^2},
 \en
where $B$ is the parity-conserving amplitude given in Eq.
(\ref{PCSigcp}) and $q^2=(p_{\bar p}+p_\pi)^2$. In \cite{CYBDNN}
we have shown that
 \be
  g_{\pi NN}(q^2) = {q^2-m_\pi^2\over
 2\sqrt{2} f_\pi  m_N}\,g_3^{np}(q^2),
 \en
where $g_3^{np}$ is one of the form factors defined by
   \be \label{f,g}
\la \bar n(p_{\bar n})|(V- A)_\mu|\bar p(p_{\bar p})\ra &=& \bar
v_{\bar n}(p_{\bar
n})\Bigg\{f_1^{np}(q^2)\gamma_\mu+i{f_2^{np}(q^2)\over 2m_N}
\sigma_{\mu\nu}q^\nu+{f_3^{np}(q^2)\over 2m_N}q_\mu \non
\\ && -
\Big[g_1^{np}(q^2)\gamma_\mu+i{g_2^{np}(q^2)\over 2m_N}
\sigma_{\mu\nu}q^\nu+{g_3^{np}(q^2)\over
2m_N}q_\mu\Big]\gamma_5\Bigg\}v_{\bar p}(p_{\bar p}),
 \en
with $q=p_{\bar n}-p_{\bar p}$. As shown in \cite{CYBaryon}, the
induced pseudoscalar form factor $g_3$ corresponds to a pion pole
contribution to the $n-p$ axial matrix element and it is related
to the form factor $g_1$ via
 \be
 g_3^{np}(t)=-{4m_N^2\over t-m_\pi^2}g_1^{np}(t).
 \en

\begin{figure}[t]
\hspace{0cm}\centerline{\psfig{figure=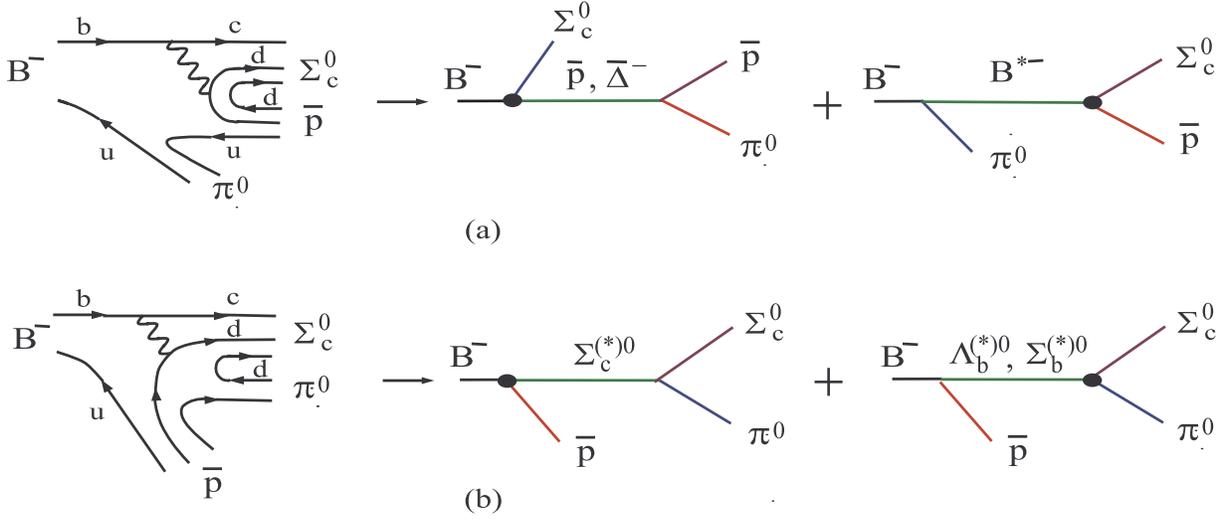,height=3.0in}}
\vspace{0.5cm}
    \caption{{\small Same as Fig. 5 except for $B^-\to\Sigma_c^0\bar
    p\pi^0$.
    }}
\end{figure}

The vector form factors $f_i(q^2)$ can be related to the nucleon's
electromagnetic form factors which are customarily described in
terms of the electric and magnetic Sachs form factors $G_E^N(t)$
and $G_M^N(t)$. A recent phenomenological fit to the experimental
data of nucleon form factors has been carried out in \cite{CHT1}
using the following parametrization:
 \be \label{GMN}
 |G_M^p(t)| &=& \left({x_1\over t^2}+{x_2\over t^3}+{x_3\over t^4}
 +{x_4\over t^5}+{x_5\over t^6}\right)\left[\ln{t\over
 Q_0^2}\right]^{-\gamma},  \non \\
|G_M^n(t)| &=& \left({y_1\over t^2}+{y_2\over
t^3}\right)\left[\ln{t\over Q_0^2}\right]^{-\gamma},
 \en
where $Q_0=\Lambda_{\rm QCD}\approx$ 300 MeV and $\gamma=2+{4\over
3\beta}=2.148$\,. For our purposes, we just need the best fit
values of $x_1$ and $y_1$
 \be
&& x_1=420.96\,{\rm GeV}^4, \qquad\quad
 y_1=236.69\,{\rm GeV}^4
 \en
extracted from the neutron data \cite{CHT2}. For the axial form
factor $g_1^{np}(t)$, we shall follow \cite{CHT3} to assume that
it has a similar expression as $G_M^n(t)$
 \be \label{g1}
 g_1^{np}(t)=\left({d_1\over t^2}+{d_2\over t^3}\right)
 \left[\ln{t\over Q_0^2}\right]^{-\gamma},
 \en
where the coefficient $d_1$ is related to $x_1$ and $y_1$ by
considering the asymptotic behavior of Sachs form factors $G_M^p$
and $G_M^n$ [see Eq. (\ref{GMN})]
 \be
 d_1={5\over 3}x_1-y_1.
 \en
For $d_2$ we shall use the value of $-2370\,{\rm GeV}^6$ obtained
by fitting to the data of $\ov B^0\to D^0p\bar p$
\cite{CYcharmful}.

Collecting all the inputs, we finally obtain
 \be
 \B(\ov B^0\to\Sigma_c^0\bar p\pi^+)^{\rm PC}_{{\bar n}-{\rm
 pole}}=1.3\times 10^{-4}.
 \en
 It is
interesting to note that this pole contribution alone is
consistent with both Belle and CLEO. The remaining pole diagrams
in Fig. 5 are more difficult to get a reliable estimate as the
strength and momentum dependence of the strong couplings is
unknown. Therefore, whether or not $\Sigma_c^0\bar p\pi^+$ is
substantially suppressed relative to $\Sigma_c^{++}\bar p\pi^-$ is
unknown. Nevertheless, as noted in passing, even if the former is
suppressed relative to the latter, it has nothing to do with color
suppression.

Likewise, we find the $\bar p$ pole diagram in Fig. 6 gives
 \be
 \B(B^-\to\Sigma_c^0\bar p\pi^0)_{{\bar p}-{\rm
 pole}}=4.8\times 10^{-3}.
 \en
This enormously large branching ratio comes from the fact that
$\Sigma_c^0\bar p\gg \Sigma_c^0\bar n$ as discussed in Sec. IIB.
At first sight, it appears that this prediction is ruled out as it
already exceeds the measurement by CLEO (see Table I). However,
the decay amplitude of Fig. 6(b) has a sign opposite to that of
Fig. 6(a) owing to the $\pi^0$ wave function $\pi^0=(\bar uu-\bar
dd)/\sqrt{2}$. Hence, there exists a destructive interference
between Fig. 6(a) and Fig. 6(b). Unfortunately, as we do not have
a reliable estimate of other pole diagrams in Fig. 6, we cannot
make a reliable prediction of the branching ratio for
$B^-\to\Sigma_c^0\bar p\pi^0$. Nevertheless, it is very
conceivable that $\Sigma_c^0\bar p\pi^0$ has a larger rate than
$\Sigma_c^0\bar p\pi^+$. Recall that the CLEO measurements imply
that $\Sigma_c^0\bar p\pi^0\gsim \Sigma_c^{++}\bar p\pi^-\gsim
\Sigma_c^0\bar p\pi^+$ \cite{CLEO02}.

\section{Conclusions}

We have studied exclusive $B$ decays to final states containing a
charmed baryon within the framework of the pole model. We first
draw some conclusions and then proceed to discuss some sources of
theoretical uncertainties.

\begin{enumerate}
\item
In the pole model, the two-body baryonic $B$ decay amplitudes are
expressed in terms of strong couplings and weak baryon-baryon
transition matrix elements. We apply the bag model to evaluate the
baryon matrix elements. Since the strong coupling for
$\Lambda_b^0\to B^-p$ is larger than that for $\Sigma_b^0\to
B^-p$, the two-body charmful decay $B^-\to\Sigma_c^0\bar p$ has a
rate larger than $\ov B^0\to\Lambda_c^+\bar p$ as the former
proceeds via the $\Lambda_b$ pole while the latter via the
$\Sigma_b$ pole. However, the relative coupling strength predicted
by the quark-antiquark creation $^3P_0$ model, namely,
$g_{\Lambda_b^0\to B^-p}=3\sqrt{3}\,g_{\Sigma_b^0\to B^-p}$ will
lead to a large rate for $B^-\to\Sigma_c^0\bar p$ that already
exceeds the present experiment limit. Likewise, the $^3P_0$
relation $g_{\Sigma_b^+\to B^-\Delta^{++}} =
2\sqrt{6}\,g_{\Sigma_b^+\to\ov B^0p}$ will lead to too large $
B^-\to p\bar\Delta^{--}$ and $B^-\to\Lambda_c^+\bar\Delta^{--}$.
Our best values for strong couplings are $|g_{\Sigma_b^+\to
B^-\Delta^{++}}|\sim 10$, $|g_{\Lambda_b^0\to B^-p}|\sim 7$ and
$|g_{\Sigma_b^0\to B^-p}|\sim 3.5$. The inconsistency of the
$^3P_0$ model's predictions with experiment may imply that the
relevant one is the $^3S_0$ model for quark pair creation.

\item The ratio of $R\equiv \Gamma(\ov B^0\to\Sigma_c^0\bar
n)/\Gamma(B^-\to\Sigma_c^0\bar p)$ also provides a nice test on
the $^3P_0$ model. While $R$ is predicted to be 1/2 in the $^3P_0$
model, it is of order only 0.01 in our case.

\item
At the quark level, $\ov B^0\to\Sigma_c^{++}\bar p\pi^-$ and
$B^-\to\Lambda_c^+\bar p\pi^-$ are expected to have similar rates
as they both receive external $W$-emission contributions. By the
same token as two-body decays,  the three-body decay $\ov
B^0\to\Sigma_c^{++}\bar p\pi^-$ receives less baryon-pole
contribution than $B^-\to\Lambda_c^+\bar p\pi^-$ at the
pole-diagram level. However, because the important charmed-meson
pole diagrams contribute constructively to the former and
destructively to the latter, $\ov B^0\to\Sigma_c^{++}\bar p\pi^-$
has a rate slightly larger than $B^-\to\Lambda_c^+\bar p\pi^-$.
\item
$B^-\to\Lambda_c^+\bar p\pi^-$ also receives the resonant
contributions from $B^-\to\Sigma_c^0\bar p$ and
$B^-\to\Lambda_c^+\bar\Delta^{--}$. The nonresonant contribution
to the branching ratio is smaller than our previous estimate
mainly because the strong coupling $g_{\Lambda_b^0\to B^-p}$
becomes smaller as implied by the data of $B^-\to\Sigma_c^0\bar
p$. We found that resonant contributions account for about one
quarter of the $B^-\to\Lambda_c^+\bar p\pi^-$ rate.

\item
The decays $\ov B^0\to\Sigma_c^0\bar p\pi^+$ and
$B^-\to\Sigma_c^0\bar p\pi^0$ that can only proceed via an
internal $W$-emission are not color suppressed. In the pole model,
it is easily seen that the weak baryon-baryon transition vertex in
the pole diagram is proportional to $(c_1-c_2)$ rather than $a_2$.
We have estimated the neutron pole contribution to $\ov
B^0\to\Sigma_c^0\bar p\pi^+$ and the proton pole contribution to
$B^-\to\Sigma_c^0\bar p\pi^0$. Due to the lack of information of
the momentum dependence of strong couplings, we cannot have a
definite prediction for the decay rates of these two modes, though
it is conceivable that $\Sigma_c^0\bar p\pi^0>\Sigma_c^0\bar
p\pi^+$.  If these two decays are found to be suppressed relative
to $\Sigma_c^{++}\bar p\pi^-$, it has nothing to do with color
suppression and must arise from some other dynamic consideration.

\end{enumerate}

The calculation of baryonic $B$ decays is rather complicated and
very much involved and hence it suffers from many possible
theoretical uncertainties. Many of them have been discussed in
detail in \cite{CYBaryon}. For the present work, we would like to
mention three uncertainties. First, the charmed meson-pole
amplitude is sensitive to the form factor $V_0^{BD_1}(0)$ which we
have taken it to be 0.37\,. Hence, a model calculation of the form
factors $V_{0,1,2}^{BD_1}(0)$ is urgent. Second, a reliable
estimate of the strong couplings for $\Sigma_c^{++}\to p
D(D^*,D_1)$ and their $q^2$ dependence is needed in order to
calculate the rate of $\ov B^0\to\Sigma_c^{++}\bar p\pi^-$ more
accurately. Third, final-state interactions may contribute sizably
to the charmful two-body baryonic $B$ decays. For example, the
color- and Cabibbo-allowed decay $\ov B^0\to D^+\pi^-$ followed by
the rescattering of $D^+\pi^-\to \Lambda_c^+\bar p$ may introduce
a significant contribution to $\ov B^0\to\Lambda_c^+\bar p$. This
deserves a further study.

\vskip 2.5cm \acknowledgments This work was supported in part by
the National Science Council of R.O.C. under Grant Nos.
NSC91-2112-M-001-038 and NSC91-2112-M-033-013.

%%%%%%%%%%%%%%%%%%%%%%%%%%%%%%%%%%%%%%%%%%%%%%%%%%%%%%%%

\end{document}